\documentclass[12pt]{amsart}

\usepackage{amssymb,latexsym,amsfonts,verbatim,amscd}

\newtheorem{thm1}{Theorem}[section]

\newtheorem{def1}[thm1]{Definition}

\newtheorem{prop1}[thm1]{Proposition}

\begin{document}

\title[Parabosonic and parafermionic algebras.]
{Parabosonic and parafermionic algebras. Graded structure and Hopf
structures}
\author[K. Kanakoglou, C. Daskaloyannis]{K. Kanakoglou, C. Daskaloyannis}
\address {K. Kanakoglou: Department of Physics, Aristotle University of Thessaloniki, Thessaloniki 54124, GREECE
 } \email{kanakoglou@hotmail.com}
\address {C. Daskaloyannis: Department of Mathematics, Aristotle University of Thessaloniki, Thessaloniki 54124, GREECE
 } \email{daskalo@math.auth.gr}

\begin{abstract}
Parabosonic $P_{B}^{(n)}$ and parafermionic $P_{F}^{(n)}$ algebras
are described as quotients of the tensor algebras of suitably
choosen vector spaces. Their (super-) Lie algebraic structure and
consequently their (super-) Hopf structure is shortly discussed. A
bosonisation-like construction is presented, which produces an
ordinary Hopf algebra $P_{B(K^{\pm})}^{(n)}$ starting from the
super Hopf algebra $P_{B}^{(n)}$.
\end{abstract}

\maketitle

\section{Introduction and Definitions}

Throughout this paper we are going to use the following notation
conventions: \\
If $x$ and $y$ are any monomials of the tensor algebra of some
k-vector space, we are going to call commutator the following
expression:
$$
[x,y] = x \otimes y - y \otimes x \equiv xy - yx
$$
and anticommutator the following expression:
$$
\{x,y\} = x \otimes y + y \otimes x \equiv xy + yx
$$
By the field $k$ we shall always mean $\mathbb{R}$ or
$\mathbb{C}$, and all tensor products will be considered over $k$
unless stated so. Finally we freely use Sweedler's notation for
the comultiplication throughout the paper.

 Let us consider the k-vector space $V_{B}$ freely generated by
 the elements: $b_{i}^{+}, b_{j}^{-}$, $i,j=1,...,n$. Let
$T(V_{B})$ denote the tensor algebra of $V_{B}$ (i.e.: the free
algebra generated by the elements of the basis). In $T(V_{B})$ we
consider the (two-sided) ideal $I_{B}$ generated by the following
elements:
$$
\begin{array}{c}
  \big[ b_{m}^{-}, \{ b_{k}^{+}, b_{l}^{-} \} \big] - 2\delta_{km}b_{l}^{-}  \\
           \\
  \big[ b_{m}^{-}, \{ b_{k}^{-}, b_{l}^{-} \} \big] \\
           \\
  \big[ b_{m}^{+}, \{ b_{k}^{-}, b_{l}^{-} \} \big] + 2\delta_{lm}b_{k}^{-} + 2\delta_{km}b_{l}^{-}\\
\end{array}
$$
and all adjoints. (Since we do not yet consider any particular
representation the term ``adjoint" has no other meaning but to
denote different (linearly independent) elements of the algebra.
We say for example that $b_{i}^{-}$ and $b_{i}^{+}$ are adjoint
elements or that $b_{i}^{+}b_{j}^{-}$ and $b_{j}^{+}b_{i}^{-}$ are
adjoint elements. But when we consider representations of such
algebras then we see that the ``physically interesting" modules
are those in which the ``adjoint" elements become adjoint
operators of a Hilbert space.) \\
 We now have the following:

\begin{def1}
We define the parabosonic algebra in $2n$ generators $P_{B}^{(n)}$
($n$ parabosons) to be the quotient algebra of the tensor algebra
of $V_{B}$ with the ideal $I_{B}$. In other words:
$$
P_{B}^{(n)} = T(V_{B}) / I_{B}
$$
\end{def1}

In a similar way we may describe the parafermionic algebra in $n$
generators: Let us consider the k-vector space $V_{F}$ freely
generated by the elements: $f_{i}^{+}, f_{j}^{-}$, $i,j=1,...,n$.
Let $T(V_{F})$ denote the tensor algebra of $V_{F}$ (i.e.: the
free algebra generated by the elements of the basis). In
$T(V_{F})$ we consider the (two-sided) ideal $I_{F}$ generated by
the following elements:
$$
\begin{array}{c}
  \big[ f_{m}^{-}, [ f_{k}^{+}, f_{l}^{-} ] \big] - 2\delta_{km}f_{l}^{-} \\
           \\
  \big[ f_{m}^{-}, [ f_{k}^{-}, f_{l}^{-} ] \big] \\
           \\
  \big[ f_{m}^{+}, [ f_{k}^{-}, f_{l}^{-} ] \big] + 2\delta_{lm}f_{k}^{-} - 2\delta_{km}f_{l}^{-}\\
\end{array}
$$
and all adjoints. \\
We get the following definition:

\begin{def1}
We define the parafermionic algebra in $2n$ generators
$P_{F}^{(n)}$ ($n$ parafermions) to be the quotient algebra of the
tensor algebra of $V_{F}$ with the ideal $I_{F}$. In other words:
$$
P_{F}^{(n)} = T(V_{F}) / I_{F}
$$
\end{def1}

Parafermionic and parabosonic algebras first appeared in the
physics literature by means of generators and relations, in the
pionnering works of Green \cite{Green} and Greenberg and Messiah
\cite{GreeMe}. Their purpose was to introduce generalizations of
the usual bosonic and fermionic algebras of quantum mechanics,
capable of leading to generalized versions of the Bose-Einstein
and Fermi-Dirac statistics (see: \cite{OhKa}).

\section{(super-)Lie and (super-)Hopf algebraic structure of $P_{B}^{(n)}$ and $P_{F}^{(n)}$}

Due to it's simpler nature, parafermionic algebras were the first
to be identified as the universal enveloping algebras (UEA) of
simple Lie algebras. This was done  almost at the same time by
S.Kamefuchi, Y.Takahashi in \cite{Kata} and by C. Ryan, E.C.G.
Sudarshan in \cite{RySu}. In fact the following proposition was
shown in the above mentioned references:

\begin{prop1}
The parafermionic algebra in $2n$ generators is isomorphic to the
universal enveloping algebra of the simple complex Lie algebra
$B_{n}$ (according to the well known classification of the simple
complex Lie algebras), i.e:
$$
P_{F}^{(n)} \cong U(B_{n})
$$
\end{prop1}

An immediate consequence of the above identification is that
parafermionic algebras are (ordinary) Hopf algebras, with the
generators $f_{i}^{\pm}$, $i=1,...,n$ being the primitive
elements. The Hopf algebraic structure of $P_{F}^{(n)}$ is
completely determined by the well known Hopf algebraic structure
of the Lie algebras, due to the above isomorphism. For convenience
we quote the relations explicitly:
$$
\begin{array}{c}
  \Delta(f_{i}^{\pm}) = f_{i}^{\pm} \otimes 1 + 1 \otimes f_{i}^{\pm} \\
      \\
   \varepsilon(f_{i}^{\pm}) = 0 \\
     \\
    S(f_{i}^{\pm}) = -f_{i}^{\pm} \\
\end{array}
$$
\\

The algebraic structure of parabosons seemed to be somewhat more
complicated. The presence of anticommutators among the (trilinear)
 relations defining $P_{B}^{(n)}$ ``breaks'' the usual (Lie)
antisymmetry and makes impossible the identification of the
parabosons with the UEA of any Lie algebra. It was in the early
'80 's that was conjectured \cite{OhKa}, that due to the mixing of
commutators and anticommutators in $P_{B}^{(n)}$ the proper
mathematical ``playground" should be some kind of Lie superalgebra
(or: $\mathbb{Z}_{2}$-graded Lie algebra). Starting in the early
'80 's, and using the recent (by that time) results in the
classification of the finite dimensional simple complex Lie
superalgebras which was obtained by Kac (see: \cite{Kac1, Kac2}),
T.D.Palev managed to identify the parabosonic algebra with the UEA
of a certain simple complex Lie superalgebra. In \cite{Pal3},
\cite{Pal5}, T.D.Palev shows the following proposition:

\begin{prop1} \label{parab}
The parabosonic algebra in $2n$ generators is isomorphic to the
universal enveloping algebra of the classical simple complex Lie
superalgebra $B(0,n)$ (according to the classification of the
simple complex Lie superalgebras given by Kac), i.e:
$$
P_{B}^{(n)} \cong U(B(0,n))
$$
\end{prop1}

Note that $B(0,n)$  in Kac's notation, is the classical simple
complex orthosymplectic Lie superalgebra denoted $osp(1,2n)$ in
the notation traditionally used by physicists until then.

The universal enveloping algebra $U(L)$ of a Lie superalgebra $L$
is not a Hopf algebra, at least in the ordinary sense. $U(L)$ is a
$\mathbb{Z}_{2}$-graded associative algebra (or: superalgebra) and
it is a super-Hopf algebra in a sense that we briefly describe:
First we consider the braided tensor product algebra $U(L)
\underline{\otimes} U(L)$, which means the vector space $U(L)
\otimes U(L)$ equipped with the associative multiplication:
$$
(a \otimes b) \cdot (c \otimes d) = (-1)^{|b||c|}ac \otimes bd
$$
for $b,c$ homogeneous elements of $U(L)$, and $\ |.| \ $ denotes
the degree of an homogeneous element (i.e.: $|b|=0$ if $b$ is an
even element and $|b|=1$ if $b$ is an odd element). Note that
$U(L) \underline{\otimes} U(L)$ is also a superalgebra or:
$\mathbb{Z}_{2}$-graded associative algebra. Then $U(L)$ is
equipped with a coproduct
$$
\Delta : U(L) \rightarrow U(L)
\underline{\otimes} U(L)
$$
which is an superalgebra homomorphism from $U(L)$ to the braided
tensor product algebra  $U(L) \underline{\otimes} U(L)$ :
$$
\Delta(ab) = \sum (-1)^{|a_{2}||b_{1}|}a_{1}b_{1} \otimes
a_{2}b_{2} = \Delta(a) \cdot \Delta(b)
$$
for any $a,b$ in $U(L)$, with $\Delta(a) = \sum a_{1} \otimes
a_{2}$, $\Delta(b) = \sum b_{1} \otimes b_{2}$, and $a_{2}$,
$b_{1}$ homogeneous.  $\Delta$ is uniquely determined by it's
value on the generators of $U(L)$ (i.e.: the basis elements of
$L$):
$$
\Delta(x) = 1 \otimes x + x \otimes 1
$$
 Similarly, $U(L)$ is equipped with an antipode $S : U(L)
\rightarrow U(L)$ which is not an algebra anti-homomorphism (as in
ordinary Hopf algebras) but a braided algebra anti-homomorphism
(or: ``twisted" anti-homomorphism) in the following sense:
$$
S(ab) = (-1)^{|a||b|}S(b)S(a)
$$
for any homogeneous $a,b \in U(L)$. \\
All the above description is equivalent to saying that $U(L)$ is a
a Hopf algebra in the braided category of
$\mathbb{CZ}_{2}$-modules, or: a braided group, where the braiding
is induced by the non-trivial quasitriangular structure of the
$\mathbb{CZ}_{2}$ Hopf algebra i.e. by the non-trivial $R$-matrix:
$$
R_{g} = \frac{1}{2}(1 \otimes 1 + 1 \otimes g + g \otimes 1 - g
\otimes g)
$$
where $1, g$ are the elements of the $\mathbb{Z}_{2}$ group which
is now written multiplicatively.

In view of the above description, an immediate consequence of
proposition \ref{parab}, is that the parabosonic algebras
$P_{B}^{(n)}$ are super-Hopf algebras, with the generators
$b_{i}^{\pm}$, $i=1,...,n$ being the primitive elements. It's
super-Hopf algebraic structure is completely determined by the
super-Hopf algebraic structure of Lie superalgebras, due to the
above mentioned isomorphism. Namely the following relations
determine completely the super-Hopf algebraic structure of
$P_{B}^{(n)}$:
$$
\begin{array}{c}
  \Delta(b_{i}^{\pm}) = 1 \otimes b_{i}^{\pm} + b_{i}^{\pm} \otimes 1 \\
    \\
  \varepsilon(b_{i}^{\pm}) = 0 \\
    \\
  S(b_{i}^{\pm}) = - b_{i}^{\pm} \\
\end{array}
$$

\section{Bosonisation: ordinary Hopf algebraic structures for parabosons}

A general scheme for ``transforming" a Hopf algebra $B$ in the
braided category ${}_{H}\mathcal{M}$ ($H$: some quasitriangular
Hopf algebra) into an ordinary one, namely the smash product Hopf
algebra: $B \rtimes H$, such that the two algebras have equivalent
module categories, has been developed during '90 's. The original
reference is \cite{Maj1} (see also \cite{Maj2, Maj3}). The
technique is called bosonisation, the term coming from physics.
This technique uses ideas developed in \cite{Ra}, \cite{Mo}. It is
also presented and applied in \cite{Fi},
\cite{FiMon}, \cite{Andru}. \\
In the special case that $B$ is some super-Hopf algebra, then: $H
= \mathbb{CZ}_{2}$, equipped with it's non-trivial quasitriangular
structure, formerly mentioned. In this case, the technique
simplifies and the ordinary Hopf algebra produced is the smash
product Hopf algebra $B \rtimes \mathbb{CZ}_{2}$. The grading in
$B$ is induced by the $\mathbb{CZ}_{2}$-action on $B$:
$$
g \vartriangleright x = (-1)^{|x|}x
$$
for $x$ homogeneous in $B$. (Note that for self-dual Hopf algebras
-such as $\mathbb{CZ}_{2}$- the notions of action and coaction
coincide).  This action is ``absorbed" in $B \rtimes
\mathbb{CZ}_{2}$, and becomes an inner automorphism:
$$
gxg = (-1)^{|x|}x
$$
where we have identified: $b \rtimes 1 \equiv b$ and $1 \rtimes g
\equiv g$ in $B \rtimes \mathbb{CZ}_{2}$. This inner automorphism
is exactly the adjoint action of $g$ on $B \rtimes
\mathbb{CZ}_{2}$ (as an ordinary Hopf algebra). The following
proposition is proved -as an example of the bosonisation
technique- in \cite{Maj2}:
\begin{prop1} \label{bosonisat}
Corresponding to every super-Hopf algebra $B$ there is an ordinary
Hopf algebra $B \rtimes \mathbb{CZ}_{2}$, it's bosonisation,
consisting of $B$ extended by adjoining an element $g$ with
relations, coproduct, counit and antipode:
$$
\begin{array}{cccc}
  g^{2} = 1 & gb = (-1)^{|b|}bg & \Delta(g) = g \otimes g & \Delta(b) = \sum b_{1}g^{|b_{2}|} \otimes b_{2} \\
                                     \\
  S(g) = g & S(b) = g^{-|b|}\underline{S}(b) & \varepsilon(g) = 1 & \varepsilon(b) = \underline{\varepsilon}(b) \\
\end{array}
$$
where $\underline{S}$ and $\underline{\varepsilon}$ denote the
original maps of the super-Hopf algebra $B$. \\
Moreover, the representations of the bosonised Hopf algebra $B
\rtimes \mathbb{CZ}_{2}$ are precisely the super-representations
of the original superalgebra $B$.
\end{prop1}
The application of the above proposition in the case of the
parabosonic algebra $P_{B}^{(n)} \cong U(B(0,n))$ is
straightforward: we immediately get it's bosonised form
$P_{B(g)}^{(n)}$ which by definition is:
$$
P_{B(g)}^{(n)} \equiv P_{B}^{(n)} \rtimes \mathbb{CZ}_{2} \cong
U(B(0,n)) \rtimes \mathbb{CZ}_{2}
$$
\\
Let us describe now a slightly different construction (see:
\cite{DaKa}), which achieves the same object: the determination of
an ordinary Hopf structure for the parabosonic algebra
$P_{B}^{(n)}$. \\
 Defining:
$$
N_{lm} = \frac{1}{2}\{b_{l}^{+},b_{m}^{-}\}
$$
we notice that these are the generators of the Lie algebra $u(n)$:
$$
[N_{kl},N_{mn}] = \delta_{lm}N_{kn} - \delta_{kn}N_{ml}
$$
We introduce now the elements:
$$
\mathcal{N} = \sum_{i=1}^{n}N_{ii} =
\frac{1}{2}\sum_{i=1}^{n}\{b_{i}^{+},b_{i}^{-}\}
$$
which are exactly the linear Casimirs of $u(n)$. \\
We can easily find that they satisfy:
$$
[\mathcal{N}, b_{i}^{\pm}] = \pm b_{i}^{\pm}
$$
We now introduce the following elements:
$$
K^{+} = \exp(i \pi \mathcal{N}) \equiv \sum_{m=0}^{\infty}
\frac{(i \pi \mathcal{N})^{m}}{m!}
$$
and:
$$
K^{-} = \exp(-i \pi \mathcal{N}) \equiv \sum_{m=0}^{\infty}
\frac{(-i \pi \mathcal{N})^{m}}{m!}
$$
Utilizing the above power series expressions and the commutation
relations formerly calculated for $\mathcal{N}$ we get after
lengthy but straightforward calculations:
$$
K^{+}K^{-} = K^{-}K^{+} = 1
$$
and also:
$$
\begin{array}{lr}
 \{K^{+},b_{i}^{\pm}\} = 0 &  \{K^{-},b_{i}^{\pm}\} = 0 \\
\end{array}
$$
We finally have the following proposition:
\begin{prop1} \label{altern}
Corresponding to the super-Hopf algebra $P_{B}^{(n)}$ there is an
ordinary Hopf algebra $P_{B(K^{\pm})}^{(n)}$, consisting of
$P_{B}^{(n)}$ extended by adjoining two elements $K^{+}$, $K^{-}$
with relations, coproduct, counit and antipode:
$$
\begin{array}{cc}
  \Delta(b_{i}^{\pm}) = b_{i}^{\pm} \otimes 1 + K^{\pm} \otimes b_{i}^{\pm} & \Delta(K^{\pm}) = K^{\pm} \otimes K^{\pm} \\
       \\
  \varepsilon(b_{i}^{\pm}) = 0 & \varepsilon(K^{\pm}) = 1 \\
                     \\
  S(b_{i}^{\pm}) = b_{i}^{\pm}K^{\mp} & S(K^{\pm}) = K^{\mp} \\
                      \\
  K^{+}K^{-} = K^{-}K^{+} = 1 & \{K^{+},b_{i}^{\pm}\} = 0 = \{K^{-},b_{i}^{\pm}\} \\
\end{array}
$$
\end{prop1}
\begin{proof}
The proposition may be proved by lengthy but straightforward
calculations. See \cite{DaKa}.
\end{proof}
The above constructed algebra $P_{B(K^{\pm})}^{(n)}$, is an
ordinary Hopf algebra in the sense that the comultiplication is
extended to the whole of $P_{B(K^{\pm})}^{(n)}$ as an algebra
homomorphism :
$$
\Delta : P_{B(K^{\pm})}^{(n)}\rightarrow P_{B(K^{\pm})}^{(n)}
\otimes P_{B(K^{\pm})}^{(n)}
$$
where $P_{B(K^{\pm})}^{(n)} \otimes P_{B(K^{\pm})}^{(n)}$ is
considered as the tensor product algebra with the usual product:
$$
(a \otimes b)(c \otimes d) = ac \otimes bd
$$
for any $a,b,c,d \in P_{B(K^{\pm})}^{(n)}$ and the antipode
extends as usual as an algebra anti-homomorphism.
\\

The relation between the above constructed ordinary Hopf algebras,
$P_{B(g)}^{(n)}$ and $P_{B(K^{\pm})}^{(n)}$ remains to be seen.
\\ \\  \\

\thanks{This research was co-funded by the European Union
in the framework of the program ``Pythagoras II", contract number
80897.

\end{document}